\documentclass[%
 reprint,
nofootinbib,
 amsmath,amssymb,
 aps,
 showkeys,
]{revtex4-2}

\usepackage{graphicx}
\usepackage{dcolumn}
\usepackage{bm}
\usepackage{bbm}

\usepackage{amsmath,amssymb,graphicx,bm}

\begin{document}
	
	\title{Analytic Cancellation of Interference Terms and Closed-Form 1-Mode Marginals in Canonical Boson Sampling}
	
	\author{Jiang Liu}
	\email{jiangliu.message@gmail.com}
	\affiliation{Independent Researcher, Scarsdale, NY 10533, USA}
	
	\date{February 28, 2026}
	
	\begin{abstract}
		Although the $k$-mode marginal distributions of Canonical Boson Sampling (CBS) are known to be computable in polynomial time, the physical mechanism driving this computational efficiency remains mathematically opaque. In this work, we provide a direct, bottom-up physical derivation of the exact 1-mode marginal distribution in CBS, computable in $\mathcal{O}(R^2)$ time, where $R$ is the total number of photons. We explicitly bridge this physical derivation with the mathematical theory of rank-1 matrix permanents, proving that multiphoton interference natively reduces to a symmetric polynomial scaled by a factorial bosonic bunching factor. Crucially, we demonstrate that our recursive combinatorial formulation circumvents the algorithmic overhead of characteristic function methods, entirely bypassing the need for polynomial interpolation or Fourier transforms. Finally, we apply this formula to identify macroscopic signatures of bunching, providing a rigorous, highly scalable metric for distinguishing genuine quantum interference from classical distinguishable-particle models using standard threshold detectors.
	\end{abstract}
	
	\maketitle
	
	\section{INTRODUCTION}
	
	The experimental validation of Canonical Boson Sampling (CBS) \cite{AA2011} remains a significant challenge in the quest for quantum advantage. While Gaussian Boson Sampling (GBS) \cite{hamilton2017} permits validation through polynomial-time low-order marginals \cite{Quesada2020}, validating large-scale CBS devices against classical spoofing models, such as the emission of distinguishable particles, has remained a critical bottleneck due to the \#P-hard nature of the ideal joint distribution.
	
	It has been established that the marginal distribution of $k$ modes in a CBS experiment can be computed deterministically in $\mathcal{O}(R^{\mathcal{O}(k)})$ time \cite{Gurvits2005}. However, this algorithmic bound relies on the evaluation of abstract characteristic functions and the permanents of artificial matrices. While this formally establishes that 1-mode marginals are computable in polynomial time, the top-down mathematical framework obscures the underlying physics. Specifically, it leaves open the question of \textit{how} the highly complex, phase-dependent multiphoton interference terms physically collapse when unobserved modes are traced out.
	
	Furthermore, from a practical standpoint, extracting the actual probability distribution from a characteristic generating function requires evaluating the permanent at multiple interpolation points and applying an Inverse Fast Fourier Transform (or solving a linear system). This introduces unnecessary numerical overhead and potential floating-point instabilities, presenting a barrier to immediate experimental implementation for large systems.
	
	In this work, we demystify this computational collapse by providing a direct, bottom-up physical derivation of the exact 1-mode marginal distribution in CBS. By applying a probability sum rule and enforcing the row-orthonormality of the transition matrix, we rigorously show that all complex interference cross-terms analytically cancel. As a result, the marginal probability depends exclusively on the absolute squares of the matrix elements of the observed mode. 
	
	Our derivation yields a highly efficient, closed-form combinatorial formula that provides the exact probabilities natively via a simple $\mathcal{O}(R^2)$ dynamic programming recurrence, entirely circumventing the exponential bottleneck of the full joint distribution and the interpolation overhead of generating functions. To demonstrate the mathematical universality of our result, we explicitly unpack Gurvits's characteristic function algorithm using the linear algebra of rank-1 permanents \cite{Gurvits2005}, proving that our physical derivation identically matches the polynomial expansion of the quantum probability generating function.
	
	This exact, scalable formulation provides a direct tool for experimental validation at arbitrarily large photon numbers. By comparing the exact single-mode marginals of indistinguishable bosons against classical distinguishable particles, we isolate the fundamental factorial weight responsible for macroscopic bosonic bunching, offering a practical validation metric measurable with standard threshold detectors.
	
	\section{MARGINAL DISTRIBUTION}
	
	Consider a CBS with $R$ single-photon sources and $M$ output modes connected by an amplitude transition matrix, $V$, whose rows are orthonormal:
	\begin{equation}
		\sum_{i=1}^M v_{i',i}v^*_{j',i} = \delta_{i',j'}. \label{eq:orthonormal}
	\end{equation}
	Let $0 \le n_i \le R$ be the number of photons in mode $i$. Let $N_R = (n_1, n_2, \cdots, n_M)$ be a specific output configuration, and let $N_{R;n_k}$ denote the configuration $N_R$ with the photon count $n_k$ in mode $k$ held fixed. The marginal probability of observing exactly $n_k$ photons in mode $k$ is given by:
	\begin{equation}
		P(n_k) = \sum_{N_{R;n_k}} P(N_{R;n_k}), \label{eq:marginal_def}
	\end{equation}
	where
	\begin{equation}
		P(N_{R;n_k}) = \frac{|\text{Perm}(\mathcal{A}(N_{R;n_k}))|^2}{\prod_{i=1}^M n_i!}, \label{eq:joint_prob}
	\end{equation}
	and $\mathcal{A}(N_{R;n_k})$ is an $R \times R$ amplitude matrix whose columns are determined by $V$ as specified by the configuration $N_{R;n_k}$.
	
	Equations (\ref{eq:marginal_def}) and (\ref{eq:joint_prob}) can be simplified significantly utilizing Eq. (\ref{eq:orthonormal}) and a fundamental sum rule of probabilities, which for $R > n_k$ is given by:
	\begin{equation}
		\sum_{N_{R;n_k}} \!\! P(N_{R;n_k}) = \sum_{N_{R-1;n_k}} \sum_{\substack{i=1 \\ i \neq k}}^M \frac{n_i + 1}{R - n_k} P(N_{R-1;n_k} + 1_i), \label{eq:sum_rule}
	\end{equation}
	where $N_{R;n_k} + 1_i$ represents a configuration that adds one additional photon to mode $i$ of $N_{R;n_k}$. The derivation of Eq. (\ref{eq:sum_rule}) is detailed in Appendix A.
	
	The significance of Eq. (\ref{eq:sum_rule}) is that the sum of probabilities over $N_{R;n_k}$ can be decomposed into a nested sum of probabilities over $N_{R-1;n_k}$ with the sequential addition of a photon to each mode of $N_{R-1;n_k}$. The inner sum over $i$ enables the direct employment of the orthonormality condition in Eq. (\ref{eq:orthonormal}).
	
	Substituting Eq. (\ref{eq:sum_rule}) into Eqs. (\ref{eq:marginal_def}) and (\ref{eq:joint_prob}), applying the Laplace permanent expansion \cite{minc1984permanents, Troyansky1996} to the rows of the added $1_i$ column, utilizing row-orthonormality, and noting that:
	\begin{equation}
		P(N_{R;n_k} + 1_i) = \frac{|\text{Perm}(\mathcal{A}(N_{R;n_k} + 1_i))|^2}{(n_i + 1) \prod_{i=1}^M n_i!}, \label{eq:prob_added}
	\end{equation}
	one finds that the $\delta_{i',j'}$ and the $v_{i',k}v^*_{j',k}$ terms reduce $\sum_i \mathcal{A}(N_{R-1;n_k} + 1_i)$ into $\mathcal{A}(N_{R-1;n_k})$ and $\mathcal{A}(N_{R-1;n_k+1})$, respectively. Through this process, one degree of freedom in the configuration summation is cleanly removed.
	
	Repeating this exact procedure until all degrees of freedom in the configuration summation are depleted, the remaining matrix and sub-matrices of $\mathcal{A}$ are left with identical columns constructed solely from the elements of the $k$-th column of the transition matrix. Because these remaining matrices are rank-1, their permanents evaluate simply to the product of their matrix elements multiplied by the factorial of the matrix dimension.
	
	Let us define the set of transition probabilities to mode $k$ as:
	\begin{equation}
		C_R := \{|v_{1,k}|^2, |v_{2,k}|^2, \dots, |v_{R,k}|^2\}. \label{eq:CR}
	\end{equation}
	For any $1 \le m \le R$, we write the set of all $m$-sized subset products as:
	\begin{equation}
		C_R^{(m)} := \Big\{ \prod_{j=1}^m |v_{i_j,k}|^2 : 1 \le i_1 < i_2 < \dots < i_m \le R \Big\}, \label{eq:CRm}
	\end{equation}
	with $C_R^{(0)} = 1$. Let $S_{R;m}$ denote the sum over all elements in $C_R^{(m)}$:
	\begin{equation}
		S_{R;m} \equiv \sum_{c \in C_R^{(m)}} c. \label{eq:SRm}
	\end{equation}
	The marginal distribution of CBS is then described by the following elegant, closed-form formula:
	\begin{equation}
		P(n_k) = \sum_{m=n_k}^R (-1)^{m-n_k} m! \binom{m}{n_k} S_{R;m}. \label{eq:marginal_final}
	\end{equation}
	
	The physical significance of this formula is that $P(n_k)$ depends exclusively on the non-zero absolute squares of the $k$-th column; all other modes are effectively decoupled. Furthermore, the combinatorial nature of $S_{R;m}$ (which represents the elementary symmetric polynomials) naturally captures the permutation symmetry required by Bose-Einstein statistics \cite{Bose1924, Einstein1924, Einstein1925}, realized strictly through the subset of photons observable in mode $k$.
	
	Computationally, the sets $C_R^{(m)}$ are recursively defined:
	\begin{equation}
		C_R^{(m)} = C_{R-1}^{(m)} \cup \{ c|v_{R,k}|^2 : c \in C_{R-1}^{(m-1)} \}. \label{eq:recursion_set}
	\end{equation}
	This recursion enables the highly efficient computation of $S_{R;m}$ and, consequently, $P(n_k)$. By initializing $S_{0;0} = 1$ (and $S_{i;j}=0$ otherwise), we can solve for $S_{i;j}$ dynamically for $1 \le i \le R$ and $1 \le j \le i$:
	\begin{equation}
		S_{i;j} = |v_{i,k}|^2 S_{i-1,j-1} + S_{i-1,j}. \label{eq:dynamic_prog}
	\end{equation}
	The number of operations required to populate this array via Eq. (\ref{eq:dynamic_prog}) is exactly $R(R-1)$. As a result, for a given transition matrix $V$, the exact marginals of CBS can be computed with a query-complexity of $\mathcal{O}(R^2)$.
	
	We verified Eq. (\ref{eq:marginal_final}) by comparing it against brute-force permanent calculations (i.e., evaluating Eqs. (\ref{eq:marginal_def}) and (\ref{eq:joint_prob})) for small numbers of $R$ in Hadamard-type interferometer models \cite{Meyer1996, Ambainis2001}, detailed in Section \ref{sec:validation}. Both methods produce identical results.
	
	\section{MARGINAL DISTRIBUTION OF DISTINGUISHABLE PHOTON SAMPLING}
	
	The marginal distributions of CBS and distinguishable photon sampling relate to one another in a remarkably simple and intuitive manner.
	
	For distinguishable photons, the marginal probability of observing exactly $n_k$ counts in mode $k$, denoted $P_d(n_k)$, is straightforward to define. Defining the row indices not included in the subset $c \in C_R^{(n_k)}$ as $i_j$ for $1 \le j \le R - n_k$, it is well known that:
	\begin{equation}
		P_d(n_k) = \sum_{c \in C_R^{(n_k)}} c \prod_{j=1}^{R-n_k} \left( 1 - |v_{i_j,k}|^2 \right). \label{eq:disting_marginal}
	\end{equation}
	Expanding this expression into subset products of matrix elements, Eq. (\ref{eq:disting_marginal}) becomes:
	\begin{equation}
		P_d(n_k) = \sum_{m=n_k}^R (-1)^{m-n_k} \binom{m}{n_k} S_{R;m}. \label{eq:disting_final}
	\end{equation}
	Comparing Eq. (\ref{eq:disting_final}) to Eq. (\ref{eq:marginal_final}), we see that the distinguishable distribution differs from the quantum bosonic distribution solely by the absence of the $m!$ multiplier in the series expansion.
	
	The absence of the $m!$ enhancement factor reflects that the transition matrix for distinguishable particles, $\mathcal{P}$, operates directly on probabilities rather than probability amplitudes \cite{Tichy2015} (i.e., $\mathcal{P}_{i',i} = |v_{i',i}|^2$), and that:
	\begin{equation}
		P_d(n_k) = \sum_{N_{R;n_k}} \frac{\text{Perm}(\mathcal{P}_S(N_{R;n_k}))}{\prod_{i=1}^M n_i!}, \label{eq:disting_perm}
	\end{equation}
	where $\mathcal{P}_S(N_{R;n_k})$ is an $R \times R$ sub-matrix built from $\mathcal{P}$ as specified by $N_{R;n_k}$. Utilizing the sum rule of probabilities from Eq. (\ref{eq:sum_rule}) and the conservation of probability $\sum_{i=1}^M |v_{i',i}|^2 = 1$, we recover Eq. (\ref{eq:disting_final}) and finally Eq. (\ref{eq:disting_marginal}). Like the quantum case, the query-complexity of computing $P_d(n_k)$ is also bounded by $\mathcal{O}(R^2)$.
	
	\section{CONNECTION TO CHARACTERISTIC FUNCTIONS AND RANK-1 PERMANENTS}
	
	The closed-form formula derived in Eq. (\ref{eq:marginal_final}) provides a clear, physical picture of how probability accumulates locally. It is highly instructive to demonstrate how this exact combinatorial structure emerges natively from the formal mathematical algorithms used in computational complexity theory.
	
	Gurvits demonstrated that the expectation value of the photon number distribution over $k$ modes can be evaluated in $\mathcal{O}(R^{\mathcal{O}(k)})$ time \cite{AA2011, Gurvits2005}. However, extracting the actual probabilities from Gurvits's expectation value requires constructing a generating function and utilizing numerical methods like polynomial interpolation or the Fast Fourier Transform (FFT).
	
	For a single mode ($k=1$), we can define a formal dummy variable $x = |c_1|^2$ to construct the Probability Generating Function (PGF) for the number of photons $n_k$ in the observed mode:
	\begin{equation}
		\text{PGF}(x) = \mathbb{E}[x^{n_k}] = \sum_{m=0}^R P(n_k = m) x^m. \label{eq:pgf_def}
	\end{equation}
	According to the characteristic function algorithm, this expectation is exactly equal to $\text{Perm}(I + V)$, where $V$ is the top-left $R \times R$ submatrix of $U^\dagger \Lambda U$, and $\Lambda$ is a diagonal matrix where the first entry is $(x - 1)$ and all others are zero.
	
	Because $\Lambda$ contains only a single non-zero element, the matrix multiplication collapses simply to $V_{i,j} = \overline{u}_{1,i}(x - 1)u_{1,j}$. Consequently, $V$ is strictly a \textit{rank-1 matrix}. The diagonal entries of $V$ correspond to the classical transition probabilities $p_i = |u_{1,i}|^2$, such that $V_{i,i} = (x - 1)p_i$.
	
	To evaluate $\text{Perm}(I + V)$, we utilize the standard linear algebra identity for the permanent of the identity matrix plus a rank-1 update, which expands as the sum of the permanents of all principal submatrices $V_S$ for subsets $S$ of size $m$:
	\begin{equation}
		\text{Perm}(I + V) = \sum_{m=0}^R \sum_{|S|=m} \text{Perm}(V_S). \label{eq:perm_expansion}
	\end{equation}
	Crucially, because $V_S$ is a rank-1 matrix, its permanent is exactly $m!$ times the product of its diagonal elements. This factorial multiplier is the exact mathematical origin of bosonic bunching. Substituting the diagonal entries yields:
	\begin{equation}
		\text{Perm}(V_S) = m! \prod_{i \in S} (x - 1)p_i = m! (x - 1)^m \prod_{i \in S} p_i. \label{eq:vs_perm}
	\end{equation}
	Summing over all subsets $S$ allows us to factor out the common terms, revealing the elementary symmetric polynomial $e_m(p_1, \dots, p_R)$, which is identical to the subset product $S_{R;m}$ defined in Eq. (\ref{eq:SRm}). The quantum Probability Generating Function thus takes the exact analytic form:
	\begin{equation}
		\text{PGF}_{\text{boson}}(x) = \sum_{m=0}^R m! (x - 1)^m S_{R;m}. \label{eq:pgf_boson}
	\end{equation}
	Extracting the coefficient of $x^{n_k}$ from Eq. (\ref{eq:pgf_boson}) mathematically recovers the exact alternating sum derived physically in Eq. (\ref{eq:marginal_final}). 
	
	This synthesis proves that the analytic cancellation of complex cross-terms observed during our physical trace operation, which arises from the orthonormality of the transition matrix, is the precise physical mechanism that restricts the system to the rank-1 permanent subspace. Furthermore, comparing Eq. (\ref{eq:pgf_boson}) to the classical PGF for distinguishable particles, $\text{PGF}_{\text{dist}}(x) = \sum_{m=0}^R (x - 1)^m S_{R;m}$, explicitly isolates the $m!$ scaling factor as the sole mathematical distinction between quantum bosons and classical particles in a single-mode observation.
	
	Most importantly, from an algorithmic perspective, our physical derivation establishes that one does not need to rely on PGF interpolation. By recursively evaluating $S_{R;m}$ via Eq. (\ref{eq:dynamic_prog}), the exact marginals can be directly populated in $\mathcal{O}(R^2)$ arithmetic operations, providing a massive computational simplification over Fourier-based extraction methods.
	
	\section{VALIDATION VIA HADAMARD BOSON SAMPLING} \label{sec:validation}
	
	This section provides the technical validation for Eq. (\ref{eq:marginal_final}). We employ a Hadamard Boson Sampling (HBS) model, where the interferometer dynamics are governed by a Hadamard-type quantum random walk \cite{Meyer1996, Ambainis2001}.
	
	The HBS model is characterized by an $R \times M$ transition matrix $V$ with a banded structure, whose rows are mutually orthogonal. Following the Cox-Ross-Rubinstein methodology \cite{Cox1979}, we represent the evolution of the quantum random walk at each time step $t$ through a balanced beam splitter by:
	\begin{equation}
		\begin{pmatrix} U_{o,t} \\ D_{o,t} \end{pmatrix} = \frac{1}{\sqrt{2}} \begin{pmatrix} 1 & -1 \\ 1 & 1 \end{pmatrix} \begin{pmatrix} U_{i,t} \\ D_{i,t} \end{pmatrix}, \label{eq:hbs_evolution}
	\end{equation}
	where subscripts $i$ and $o$ represent ``in'' and ``out'' respectively.
	
\begin{figure}[h!]
		\centering
		\includegraphics[scale=0.9]{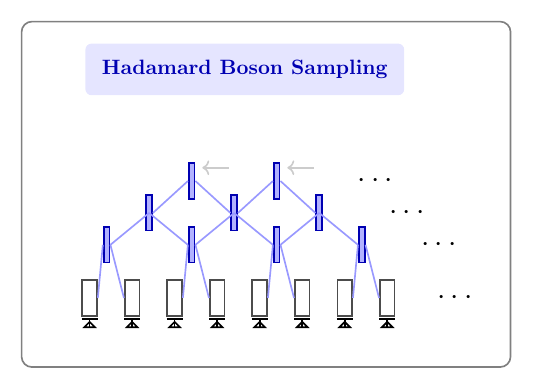}
		\caption{\label{fig:HBS}
				Schematics of Hadamard Boson Sampling for interference depth $T = 3$. Photons from the sources enter an interferometer network consisting of beam splitters arranged in a tree-like configuration implementing the Hadamard transformation, and are subsequently measured by photon detectors.
		}
	\end{figure}

	The number of layers of beam splitters, $T$, determines the depth of interference. For $T=3$ and initial conditions $U_{i,1} = 1, D_{i,1} = 0$, the single-photon amplitude vector is $v_3 = 2^{-3/2}(1, -1, 0, 2, 1, 1)$, where the zero arises from destructive Mach-Zehnder interference \cite{Mach1892, Zehnder1891}. For $R$ photons, the transition matrix $V_T$ has $R$ rows constructed by shifting $v_T$ by two columns for each subsequent row, e.g.,
	\begin{equation}
		V_3 = \left(\frac{1}{\sqrt{2}}\right)^3 \begin{pmatrix} 1 & -1 & 0 & 2 & 1 & 1 & & \\ & & 1 & -1 & 0 & 2 & 1 & 1 & \\ & & & \ddots & \ddots & \ddots & \ddots & \ddots \end{pmatrix}. \label{eq:v3_matrix}
	\end{equation}
	The schematics of the model is illustrated in Fig. \ref{fig:HBS}.

	A defining feature of the HBS model is the 2-periodicity observed in the marginal distributions. This property arises because the sets of non-zero elements in the columns of $V$ are identical for indices differing by two for modes not near the boundaries. For bulk indices where $2T - 2 < i < j < 2R + 1$, this shift-invariance ensures:
	\begin{equation}
		P(n_i) = P(n_{i+2}). \label{eq:periodicity}
	\end{equation}
	Table \ref{tab:table1} presents the full marginal distributions for $T = 3$. Our analytical results match brute-force calculations exactly for small system sizes ($R, T \in [3, 8]$), where brute-force calculation remains practical.
	
	While Table \ref{tab:table1} explores the full distribution, it is instructive to note the theoretical behavior at the extreme tail. In the limit where all $R$ photons bunch into a single observed mode ($n_k = R$), the summation in Eq. (\ref{eq:marginal_final}) collapses to a single term, yielding the exact analytic relation $P(R) = R! P_d(R)$. This massive divergence represents the ultimate theoretical manifestation of macroscopic bosonic bunching. However, despite this stark $R!$ enhancement, the underlying classical probability $P_d(R)$ scales as $\sim (1/M)^R$. Consequently, the absolute probabilities $P(R)$ and $P_d(R)$ in this high-photon regime remain vanishingly small. Resolving these rare multiphoton events with statistical significance would require prohibitively long measurement times and highly efficient photon-number-resolving (PNR) detectors, rendering the tail of the distribution largely impractical for near-term experimental validation.
	
	\begin{table*}[t]
		\caption{
			Marginal distribution of HBS for $T = 3$ and $R \ge T$, with $M = 2(R + T - 1)$. Bulk modes ($2i - 1, 2i$) exhibit 2-periodicity for $3 \le i \le R$. Probabilities for distinguishable photons $P_d(n_j)$ are provided in parentheses.}
		\label{tab:table1}
		\begin{ruledtabular}
			\begin{tabular}{cccccccc}
				\hline \hline
	  Count $n_k$ & $k \in \{1, 2, 3\}$ & $k = 4$     & $k = 2i - 1$  & $k = 2i$        & $k = M - 3$ & $k = M - 2$ & $k \in \{M - 1, M\}$ \\
				\hline
				0 & 7/8 (7/8)           & 8/16 (7/16) & 50/64 (49/64) & 62/128 (49/128) & 7/8 (7/8)   & 8/16 (7/16) & 7/8 (7/8) \\
				1 & 1/8 (1/8)           & 6/16 (8/16) & 12/64 (14/64) & 42/128 (63/128) & 1/8 (1/8)   & 6/16 (8/16) & 1/8 (1/8) \\
				2 & 0 (0)               & 2/16 (1/16) & 2/64 (1/64)   & 18/128 (15/128) & 0 (0)       & 2/16 (1/16) & 0 (0) \\
				3 & 0 (0)               & 0 (0)       & 0 (0)         & 6/128 (1/128)   & 0 (0)       & 0 (0)       & 0 (0) \\
				\hline \hline
			\end{tabular}
		\end{ruledtabular}
	\end{table*}
	
	\begin{table*}[t]
		\caption{The marginal distributions $P(n_k)$ in HBS for $n_k = 0, 1$ at mode $k$. $T$ is the number of beam-splitter layers in the interferometer. For each $T$ the number of photons, $R \ge T$, is arbitrary. Bulk modes ($2i - 1, 2i$) exhibit 2-periodicity for $3 \le i \le R$. Probabilities for distinguishable photons $P_d(n_j)$ are provided for comparison. In general, we have $P(0) > P_d(0)$ and $P_d(1) > P(1)$.}
		\label{tab:table2}
		\begin{ruledtabular}
			\begin{tabular}{c|cccc|cccc}
				\hline \hline
				& \multicolumn{4}{c|}{$k = 2i - 1$} & \multicolumn{4}{c}{$k = 2i$} \\
				\hline
				$T$ & $P(0)$ & $P(1)$ & $P_d(0)$ & $P_d(1)$ & $P(0)$ & $P(1)$ & $P_d(0)$ & $P_d(1)$ \\
				\hline
				3   & 0.78 & 0.19 & 0.77 & 0.22 & 0.48 & 0.33 & 0.38 & 0.49 \\
				4   & 0.79 & 0.17 & 0.77 & 0.21 & 0.45 & 0.39 & 0.36 & 0.54 \\
				5   & 0.68 & 0.23 & 0.63 & 0.31 & 0.50 & 0.44 & 0.47 & 0.50 \\
				6   & 0.68 & 0.23 & 0.63 & 0.31 & 0.57 & 0.32 & 0.51 & 0.42 \\
				7   & 0.76 & 0.20 & 0.74 & 0.24 & 0.55 & 0.27 & 0.45 & 0.40 \\
				8   & 0.77 & 0.19 & 0.75 & 0.23 & 0.55 & 0.27 & 0.44 & 0.41 \\
				9   & 0.70 & 0.22 & 0.65 & 0.29 & 0.57 & 0.30 & 0.50 & 0.42 \\
				10  & 0.70 & 0.22 & 0.65 & 0.29 & 0.56 & 0.32 & 0.50 & 0.42 \\
				20  & 0.76 & 0.19 & 0.73 & 0.23 & 0.57 & 0.26 & 0.48 & 0.38 \\
				30  & 0.72 & 0.21 & 0.67 & 0.27 & 0.60 & 0.25 & 0.52 & 0.36 \\
				50  & 0.72 & 0.20 & 0.68 & 0.26 & 0.61 & 0.25 & 0.53 & 0.35 \\
				100 & 0.75 & 0.19 & 0.72 & 0.24 & 0.59 & 0.25 & 0.51 & 0.35 \\
				150 & 0.73 & 0.20 & 0.69 & 0.26 & 0.61 & 0.24 & 0.53 & 0.34 \\
				\hline \hline
			\end{tabular}
		\end{ruledtabular}
	\end{table*}
	
	Crucially, macroscopic signatures of quantumness can instead be rigorously verified using standard threshold detectors, which simply distinguish between ``no-click'' (vacuum) and ``click'' (one or more photons) events. Table \ref{tab:table2} displays the marginals of zero and one-click for $3 \le T \le 150$ in modes with full bandwidth. As demonstrated, there is a persistent, noticeable difference between the exact quantum vacuum probability $P(0)$ and the classical distinguishable counterpart $P_d(0)$, even as the system scales to large photon numbers. Because indistinguishable bosons exhibit a strong tendency to bunch together into the same modes, they leave a proportionally larger number of output modes completely empty, universally resulting in $P(0) > P_d(0)$.
	
	Furthermore, the single-photon probabilities provide a secondary, complementary signature of this bunching behavior. Not only is the probability of finding an isolated single photon suppressed relative to the classical case ($P(1) < P_d(1)$), but the quantum distribution also frequently exhibits a stark internal inversion where $P(1) < P(0)$. This phenomenon serves as a robust, easily measurable hallmark of genuine multiphoton quantum interference.
	
	In models with reflection mirrors at the boundaries, the transition matrix $V$ is modified at the edges. However, because marginals are local properties dependent solely on the matrix elements of the observed modes, results for constrained and unconstrained HBS are identical for configurations removed from the interferometer boundaries.
	
	\section{CONCLUSION}
	
	We have obtained an exact, verifiable, and polynomially efficient closed-form formula for the 1-mode marginal distribution in the CBS model. By applying probability sum rules and the row-orthogonality of the transition matrix, we physically demonstrated how the unobserved multi-particle interference cross-terms analytically annihilate each other, demystifying the computational efficiency of low-order CBS marginals.
	
	A notable feature of our result is its combinatorial clarity. We proved that the marginals are entirely determined by the subset products of classical transition probabilities (the elementary symmetric polynomials) that reach the observation mode. Furthermore, by bridging our physical derivation with generating functions and rank-1 permanents, we explicitly isolated the $m!$ factorial weight as the exact mathematical mechanism separating distinguishable and indistinguishable particles. 
	
	From a computational standpoint, our direct $\mathcal{O}(R^2)$ recursive formulation provides a significant practical advantage over abstract generating function algorithms, completely circumventing the algorithmic overhead and numerical instabilities associated with polynomial interpolation or Fourier transforms. 
	
	The strict $\mathcal{O}(R^2)$ efficiency of our formula enables the practical, exact validation of CBS experiments at arbitrarily large photon populations. By comparing empirical click/no-click statistics against our exact theoretical distributions, experimentalists can rigorously verify macroscopic bosonic bunching without requiring intractable full-distribution simulations or complex photon-number-resolving arrays. While the present work focuses on the single-photon input regime, the framework extends naturally to multi-photon sources, the generalized formula of which will be presented in a forthcoming publication.
	
	\begin{acknowledgments}
		Any views expressed are mine as an individual and not as a representative speaking for or on behalf of my employer, nor do they represent my employer's positions, strategies or opinions. This research did not receive any grant from funding agencies in the public, commercial, or not-for-profit sectors. The author acknowledges the use of a large language model (Google Gemini) for assistance in mathematical literature review and algebraic synthesis regarding the rank-1 permanent expansion.
	\end{acknowledgments}
	
	\appendix
	
	\section{The Sum Rule for Boson Sampling Probabilities} \label{app:sum_rule}
	
	The derivation of our marginal distribution formula relies on a fundamental sum rule. This rule decomposes the total probability into a nested sum of probabilities, enabling us to apply the row-normality of the transition matrix directly. This rule holds for both distinguishable and indistinguishable boson sampling where the measurement is the photon number count.
	
	Let $P(N_R)$ be the probability of an $M$-mode configuration:
	\begin{equation}
		N_R = (n_1, \dots, n_k, \dots, n_M) \label{eq:app_A1}
	\end{equation}
	with $\sum_k n_k = R$. Let $N_R + 1_k$ be a configuration:
	\begin{equation}
		N_R + 1_k = (n_1, \dots, n_k + 1, \dots, n_M), \label{eq:app_A2}
	\end{equation}
	which differs from $N_R$ by having an additional photon in mode $k$. The recursive sum rule of probabilities is:
	\begin{equation}
		\sum_{N_{R+1}} P(N_{R+1}) = \sum_{k=1}^M \sum_{N_R} \frac{n_k + 1}{R + 1} P(N_R + 1_k), \label{eq:app_A3}
	\end{equation}
	where $n_k$ is an element of $N_R$.
	
	Consider $M = 3$ as an example. The configurations of $N_2$ and $N_1$ are:
	\begin{equation}
		(2, 0, 0), (1, 1, 0), (1, 0, 1), (0, 2, 0), (0, 1, 1), (0, 0, 2), \label{eq:app_A4}
	\end{equation}
	and
	\begin{equation}
		(1, 0, 0), (0, 1, 0), (0, 0, 1), \label{eq:app_A5}
	\end{equation}
	respectively. Adding one boson to each mode of $N_1$, we have the configurations:
	\begin{align}
		(1, 0, 0) &\to (2, 0, 0), (1, 1, 0), (1, 0, 1), \nonumber \\
		(0, 1, 0) &\to (1, 1, 0), (0, 2, 0), (0, 1, 1), \label{eq:app_A6} \\
		(0, 0, 1) &\to (1, 0, 1), (0, 1, 1), (0, 0, 2). \nonumber
	\end{align}
	Summing over the probabilities of these configurations with their respective weights---which from Eq. (\ref{eq:app_A3}) are $1$ for configurations $(2, 0, 0), (0, 2, 0), (0, 0, 2)$ and $1/2$ for the rest---we recover the sum of probabilities of the configurations of $N_2$.
	
	To prove Eq. (\ref{eq:app_A3}), let us define the configuration sum using the accumulative partial-sum indices:
	\begin{equation}
		n'_k = \sum_{j=k}^M n_j. \label{eq:app_A7}
	\end{equation}
	The sum over $N_R$ is an iterative accumulation:
	\begin{equation}
		\sum_{N_R} = \sum_{n'_2=0}^R \cdots \sum_{n'_{M-1}=0}^{n'_{M-2}} \sum_{n'_M=0}^{n'_{M-1}}. \label{eq:app_A8}
	\end{equation}
	Consider the $k = M - 1$ and $k = M$ terms, summing over the $M$-th mode index $n'_M$:
	\begin{align}
		S \equiv& \sum_{n'_M=0}^{n'_{M-1}} \frac{n'_{M-1} + 1 - n'_M}{R + 1} P(\dots, n'_{M-1} + 1 - n'_M, n'_M) \nonumber \\
		&+ \sum_{n'_M=0}^{n'_{M-1}} \frac{n'_M + 1}{R + 1} P(\dots, n'_{M-1} - n'_M, n'_M + 1). \label{eq:app_A9}
	\end{align}
	By shifting the index of the second term $m = n'_M + 1$, we observe that the terms ``telescope'', allowing the summation boundary to extend from $n'_{M-1}$ to $n'_{M-1} + 1$, and that $S$ is reduced to a single term:
	\begin{equation}
		S = \sum_{n'_M=0}^{n'_{M-1}+1} \frac{n'_{M-1} + 1 - n'_M}{R + 1} P(\dots, n'_{M-1} + 1 - n'_M, n'_M). \label{eq:app_A10}
	\end{equation}
	Repeating this procedure for $k = M -1, \dots, 2$, we obtain Eq. (\ref{eq:app_A3}).
	
	Keeping $n_k$ fixed and populating $R-n_k$ bosons in $M-1$ modes, following the exact same process yields Eq. (\ref{eq:sum_rule}).
	
	\section{The Marginals} \label{app:marginals}
	
	From Eqs. (\ref{eq:marginal_def}) to (\ref{eq:sum_rule}), we have:
	\begin{equation}
		P(n_i) = \sum_{N_{R-1;n_i}} \sum_{\substack{k=1 \\ k \neq i}}^M \frac{|\text{Perm}(\mathcal{A}(N_{R-1;n_i} + 1_k))|^2}{(R - n_i)\prod_{l=1}^M n_l!}. \label{eq:app_B1}
	\end{equation}
	Let $\mathcal{A}_{i'}(N_{R-1;n_i})$ be a sub-matrix of $\mathcal{A}(N_{R-1;n_i} + 1_k)$ obtained by removing the $i'$-th row and a column associated with mode $k$. Applying the Laplace permanent expansion \cite{minc1984permanents, Troyansky1996}:
	\begin{equation}
		\text{Perm}(\mathcal{A}(N_{R-1;n_i} + 1_k)) = \sum_{i'=1}^R v_{i',k}\text{Perm}(\mathcal{A}_{i'}(N_{R-1;n_i})). \label{eq:app_B2}
	\end{equation}
	Substituting Eq. (\ref{eq:app_B2}) into Eq. (\ref{eq:app_B1}) and applying the row-orthonormal property of the transition matrix in the form:
	\begin{equation}
		\sum_{\substack{k=1 \\ k \neq i}}^M v_{i',k}v^*_{j',k} = \delta_{i',j'} - v_{i',i}v^*_{j',i}, \label{eq:app_B3}
	\end{equation}
	we obtain:
	\begin{align}
		P(n_i) &= \frac{1}{R - n_i} \sum_{N_{R-1;n_i}} \sum_{c \in C_{R-1}} \Big|\mathcal{A}_c(N_{R-1;n_i})\Big|^2 \nonumber \\
		&\quad - \frac{n_i + 1}{R - n_i} \sum_{N_{R-1;n_i+1}} \Big|\mathcal{A}(N_{R-1;n_i+1})\Big|^2, \label{eq:app_B4}
	\end{align}
	where $C_{R-1}$ (with $C_0 = 1$) is a set of $(R-1)$-combinations of row indices of $V$, and $\mathcal{A}_c(N_{R-1;n_i})$ is a sub-matrix of $\mathcal{A}(N_{R-1;n_i} + 1_k)$ with one row and one column removed as in Eq. (\ref{eq:app_B2}). The first term of Eq. (\ref{eq:app_B4}) arises from the $\delta_{i',j'}$ term of Eq. (\ref{eq:app_B3}), and the second from the $v_{i',i}v^*_{j',i}$ term.
	
	Repeating the same process until all columns other than the $i$-th column are depleted:
	\begin{align}
		P(n_i) &= \sum_{k=0}^{R-n_i} \frac{(-1)^k (R - n_i - k)!}{(R - n_i)!} \frac{(n_i + k)!}{n_i!} \binom{R - n_i}{k} \nonumber \\
		&\quad \cdot \sum_{c \in C_{R-k}} \Big|\mathcal{A}_c(N_{n_i+k;n_i+k})\Big|^2 \nonumber \\
		&= \sum_{k=0}^{R-n_i} (-1)^k \binom{n_i + k}{n_i} \sum_{c \in C_{R-k}} \Big|\mathcal{A}_c(N_{n_i+k;n_i+k})\Big|^2. \label{eq:app_B5}
	\end{align}
	The profound significance of Eq. (\ref{eq:app_B5}) is that all remaining matrices have identical columns composed strictly of the elements of the $i$-th column of the transition matrix. The permanents of these rank-1 matrices are trivial to compute:
	\begin{equation}
		\mathcal{A}_c(N_{n_i+k;n_i+k}) = \sqrt{(n_i + k)!} \prod_{i'=1}^{n_i+k} v_{c_{i'},i}. \label{eq:app_B6}
	\end{equation}
	Combining Eqs. (\ref{eq:app_B5}) and (\ref{eq:app_B6}) immediately yields our closed-form formula for marginals described by Eq. (\ref{eq:marginal_final}).
	
	Finally, let us verify that $P(n_i)$ is correctly normalized:
	\begin{equation}
		\sum_{n_i=0}^R P(n_i) = 1. \label{eq:app_B7}
	\end{equation}
	For clarity, let us define:
	\begin{equation}
		G_m = \begin{cases} 1 & m = 0 \\ (m!)^2 \Big|\prod_{k=1}^m v_{c_k,i}\Big|^2 & m > 0 \end{cases} \label{eq:app_B8}
	\end{equation}
	It follows that:
	\begin{align}
		\sum_{n_i=0}^R P(n_i) &= \sum_{n_i=0}^R \sum_{m=0}^{R-n_i} (-1)^m \frac{1}{m!n_i!} G_{m+n_i} \nonumber \\
		&= \sum_{m=0}^R \sum_{n_i=0}^m \frac{(-1)^{m-n_i}}{(m - n_i)!n_i!} G_m \nonumber \\
		&= \sum_{m=0}^R \frac{(-1)^m}{m!} G_m \sum_{n_i=0}^m (-1)^{n_i} \binom{m}{n_i} \nonumber \\
		&= G_0 \nonumber \\
		&= 1. \label{eq:app_B9}
	\end{align}
	
	\bibliography{BosonSampling1}
	
\end{document}